\documentclass[draft,preprint,pre,amsmath,showpacs,nofootinbib,endfloats]{revtex4}
\usepackage{bm}
\newcommand{\bfvarphi}{\bm{\varphi}}
\newcommand{\bfp}{\bm{p}}
\newcommand{\bfw}{\bm{w}}
\newcommand{\bfx}{\bm{x}}
\newcommand{\bfy}{\bm{y}}
\newcommand{\bfl}{\bm{1}}

\begin{document}

\title{Macrostate Data Clustering}
\author{Daniel Korenblum}
\altaffiliation[Current address: ] {Gene Network Sciences, Ithaca, NY 14850}
\author{David Shalloway}
\email{dis2@cornell.edu}
\affiliation{Biophysics Program\\Dept.\ of Molecular Biology and Genetics\\Cornell University\\Ithaca, NY 14853}

\begin{abstract}
We develop an effective non-hierarchical data clustering method using an analogy to the dynamic coarse-graining of a
stochastic system.  Analyzing the eigensystem of an inter-item transition matrix identifies fuzzy clusters corresponding to the metastable macroscopic states (macrostates) of a diffusive system.  A novel ``minimum uncertainty criterion'' determines the linear
transformation from eigenvectors to cluster-defining window functions. Eigenspectrum gap and cluster certainty conditions
identify the proper number of clusters. The physically-motivated fuzzy representation and associated uncertainty analysis distinguishes macrostate clustering from spectral partitioning methods. Macrostate data clustering solves a variety of test cases that challenge other methods.
\end{abstract}

\maketitle

\section{Introduction}

Finding subgroups or \emph{clusters} within large sets of \emph{items} is a problem that occurs in many contexts from
astronomy to integrated chip design and pattern recognition (\cite{Jain:81,Mirkin:96,Everitt:01,Szallasi:01} for reviews). DNA
microarray gene expression analysis and bioinformatic sequence comparisons are recent and important applications in
molecular biology \cite{Altman:01,Szallasi:01}.

The clustering problem may be posed in two ways: In the first case (e.g., DNA microarrays), $N_M$ measurements are made on each of the $N$ items.  The $N \times N_M$ measurement matrix $X$ is then used in a problem-specific manner to compute a symmetric $N \times N$ \emph{dissimilarity} matrix $D$.  Each off-diagonal element $D_{ij}$ provides an inverse indicator of the correlations between the measurements of items $i$ and $j$.  A straightforward method is to set
\begin{equation}
\label{D}
D_{ij} = \left [ \sum_{a,b=1}^{N_M} (X_{ia}-X_{ja}) g_{ab} (X_{ib}-X_{jb})\right]^{1/2} \;,
\end{equation}
where $g$ is a problem-specific Euclidean metric tensor.  This allows preconditioning of the scales of the different measurements and, by using non-diagonal $g$, adjustment for measurement correlations (particularly important if some measurements are replicates).  Statistical noise and complexity can be reduced by using singular value decomposition to pre-identify principal components of $X$ that span much of the variation in the measurement space.  This facilitates identification of clusters ``by eye'' or with various heuristics (e.g., \cite{Drineas:99,Alter:00,Kannan:01}).

In the second case (e.g., pairwise gene sequence comparisons), the primary data are measures of dissimilarities between pairs of items:  In this case $D$ is defined, but there is no measurement matrix $X$ and the elements of $D$ may not satisfy the triangle inequality.

Early clustering methods were ``hierarchical,'' generating open binary trees which can (depending on the selected cross-section) dissect the items into any number of clusters between 1 and $N$. In these methods, the choice of the optimal number of clusters is an independent problem \cite{Everitt:01,Milligan:85,Gordon:98}.  ``Agglomerative'' hierarchical methods iteratively join items together to form a decreasing number of larger clusters; ``divisive'' hierarchical methods use successive subdivision to generate an increasing number of smaller clusters.    While agglomerative methods can be inexpensive, they usually use only local and not global information, which weakens performance \cite{Everitt:01}.  While divisive methods can use global information, they can have high complexity in $N$ and thus can be too expensive for large problems. A weakness of both types of hierarchical methods is that they cannot repair defects from previous stages of analysis.

Some clustering methods are based on analogies to physical systems in which macroscopic structure emerges from microscopically-defined interactions.  A number have used analogies to statistical mechanical phase transitions \cite{Rose:90,Blatt:96,Wiseman:98,Kullmann:00,Giada:01}, while others have used chaotic \cite{Angelini:00} or quantum mechanical \cite{Horn:02} systems as analogs. Most of these have the advantage of being ``fuzzy''---in addition to assigning items to clusters, they provide a continuous measure of the probability or strength of the assignment of each item.

Clustering can also be performed by objective function optimization.  If there is an \emph{a priori} model for the structure of the clusters in the measurement space (e.g., as a collection of Gaussians), then a corresponding parametric objective function can be used.  Otherwise a non-parametric objective function may be useful. For example, graph theory clustering methods treat the items as nodes of a graph whose interconnecting edges have ``affinities'' or ``weights'' determined from $D$ (\cite{Seary:95,Weiss:99}, for review). They typically use ``min-cut'' or ``normalized-cut'' objective functions and search for the (sometimes ``balanced'') graph partitioning that minimizes the (sometimes normalized) sum of the weights of the cut edges.  ``Spectral graph theory'' \cite{Chung:97} methods use the eigenvectors of the affinity matrix (or the closely related generalized Laplacian matrix) to assist the process.  Spectral bipartitioning (\cite{Spielman:96}, for history and review), which uses one eigenvector, can be applied recursively for hierarchical dissection \cite{Barnard:94}; and the development of non-hierarchical methods for the concurrent use of multiple eigenvectors is an active topic of research (\cite{Alpert:99,Weiss:99}, for reviews).

We present here a novel, non-hierarchical, fuzzy clustering method which uses an analogy between data clustering and the coarse-graining of a stochastic dynamical system. The items are regarded as microstates that interact via a dynamical \emph{transition matrix} $\Gamma$, which is derived from $D$. Clusters are identified as the slowly-relaxing metastable macroscopic states (macrostates) of the system.  These are computed by concurrently using multiple eigenvectors of $\Gamma$ in the same way that macrostates of a continuous diffusive system are identified from the eigenfunctions of the Smoluchowski operator \cite{Shalloway:96}. The number of clusters is algorithmically determined by the spectral properties and cluster overlap criteria.  We demonstrate that the method can solve difficult problems, including ones in which the clusters are irregularly shaped and separated by tortuous boundaries.

\section{Method}

\subsection{Macrostates and stochastic coarse-graining; a brief overview}
Coarse-graining is used in nonequilibrium statistical physics to reduce the dimensionality and  complexity of the dynamical description \cite{Kubo:85}.  In the usual situation, the system is initially described microscopically by a fine-grained first-order equation specified over a configuration space of microscopic states (\emph{microstates}).  Microscopic degrees-of-freedom corresponding to very rapid motions are removed by (possibly non-linear) projection.  This generates a coarse-grained master equation with fewer degrees-of-freedom that describes the slower dynamics of the system's macrostates. Each macrostate corresponds to a subregion of configuration space that has been projected onto one value of the macroscopic parameters.  For example, to describe Brownian motion of pollen in water, the (fast) water molecule degrees-of-freedom are projected out leaving only the (slow) coordinate of the pollen to parameterize the macrostates. In this example the macrostates are continuously parameterized, but they may also be discrete.  For example, to describe chemical reactions, each macrostate is a chemical state, a subregion of conformation space which includes all vibrations, translations, and rotations of a specific metastable, covalently-bonded  arrangement of atoms.

Coarse-graining projections are not arbitrary:  the utility of the resultant macroscopic description depends upon the existence of a sufficient disparity between $\tau^{\rm local}$, the time-scale of the fast (projected-out) motions (which generate ergodicity within the macrostate), and $\tau^{\rm global}$, the time-scale of the remaining slow motions (which are required for ergodicity between macrostates).  Appropriate projections can sometimes be chosen heuristically when the disparity between $\tau^{\rm local}$ and $\tau^{\rm global}$ is large and subjectively obvious.  When this is not so, projections into discrete macrostates can be selected by analyzing the eigenspectrum of the microscopic stochastic dynamics.  This procedure is described in detail in Refs.\ \onlinecite{Shalloway:96} and \onlinecite{Ulitsky:98}.  We summarize the salient points here.

Consider the example illustrated in Fig.\ \ref{fig:heuristic}A of a thermal ensemble of systems having microscopic parameter $x$ and
potential energy $V(x)$.  The bimodal equilibrium probability density is given by the Gibbs-Boltzmann distribution
$p^{\rm eq}(x) \propto \exp[-\beta V(x)]$, where $\beta$ is the inverse temperature in inverse energy units.  If system dynamics
are overdamped (i.e., diffusive), then the nonequilibrium probability distribution $p(x;t)$ evolves in time according to the
first-order Smoluchowski equation
\begin{equation}
\label{continuous_time}
\frac{\partial p(x;t)}{\partial t} = \int \Gamma(x,x') \, p(x';t) \, dx' \;,
\end{equation}
where $\Gamma$ is the kernel of an operator determined by $V$, the temperature, and the diffusion constant.  Once the eigenfunctions and eigenvalues of $\Gamma$ have been determined, the general solution to Eq.\ \eqref{continuous_time} can be expanded as
\begin{equation}
\label{eigenfunction_expansion}
p(x;t) = \sum_{n=0}^\infty c_n \, e^{- \gamma_n t} \, \varphi_n(x) \;,
\end{equation}
where the eigenvalues and right-eigenfunctions of  $\Gamma$ are $-\gamma_n$ and $\varphi_n(x)$, respectively, and the expansion coefficients $c_n$ are determined by the initial conditions $p(x;0)$.  (Without loss of generality we normalize $\varphi_0$ so that $c_0 = 1$, and assume that eigenfunctions are ordered according to the magnitudes of their eigenvalues.)

We always have $\gamma_0=0$ and $\varphi_0(x) = p^{\rm eq}(x)$, corresponding to the stability of the Gibbs-Boltzmann distribution.  The other $\gamma_n$ are non-negative and determine the rates of relaxation towards this equilibrium state.  While $\varphi_0$ is non-negative everywhere, the other eigenfunctions take both positive and negative values, and the exponential decays of their amplitudes generate  probability fluxes.  For illustration, Fig.\ \ref{fig:heuristic}A, panel b displays (for a selected temperature) $\varphi_0$ and the ``slow'' right-eigenfunction $\varphi_1$. If $c_1 >0$, $p(x;0)$ will have a probability excess (relative to $p^{\rm eq}$) for $x<0$ and a deficiency for $x>0$.  These overall deviations from equilibrium will decay away as $\exp(- \gamma_1 t) \to 0$.  The ``fast'' eigenfunctions $\varphi_{n>1}$ will have more nodes than $\varphi_1$ and their more rapid decays will transport probability over smaller regions. 

Sufficiently large potential energy barriers can separate configuration space into localized, dynamically metastable macrostate regions, each having the property that $\tau^{\rm local}$, the time scale for relaxation of $p(x;t)$ towards $p^{\rm eq}(x)$ within the region, is much less than $\tau^{\rm global}$, the time-scale for probability to enter or leave the region. $\tau^{\rm local}$ and $\tau^{\rm global}$ are determined by the eigenvalues, and a disparity between them  will correspond to a gap in the eigenspectrum. If there are $m$ macrostates, a gap will occur between $\gamma_{m-1}$ and $\gamma_m$: there will be $m$ slow modes generating  inter-macrostate probability equilibration, and the remaining fast modes will generate intra-macrostate relaxations.

For example, in Fig.\ \ref{fig:heuristic}A the energy barrier centered at $x=0$ separates configuration space into two macrostates $a$ and $b$ (roughly containing the regions $x<0$ or $x>0$, respectively).  Correspondingly, there is a spectral gap $\gamma_1 \ll \gamma_2$; so $m=2$.  $\gamma_1$ is the rate of the slow transfer of probability between $a$ and $b$ that is generated by the slow decay of the amplitude of $\varphi_1$.  Thus, $\tau^{\rm global} \sim \gamma_1^{-1}$.  The larger values of the $\gamma_{n >1}$ correspond to the fast decays of the more-rapidly varying $\varphi_{n >1}$, corresponding to intra-macrostate probability relaxations.  The slowest of these rates, $\gamma_2$, determines the time needed for local equilibration.  Thus, $\tau^{\rm local} \sim \gamma_2^{-1}$.

In this simple case, it is tempting to ``crisply'' define the macrostates by inspection as the regions $x>0$ and $x<0$. However, this is inapt for two reasons: (1) A sharp boundary at $x=0$ introduces high-frequency dynamical modes and thus is incompatible with a consistent low-frequency dynamical description; and (2) Subjective inspection and barrier identification are not possible in multidimensional problems.  Instead, we use this example to show how the correct ``fuzzy'' macrostates can be identified (without subjective inspection) by a generalizable algorithm:

The starting point is the recognition of the spectral gap $\gamma_1 \ll \gamma_2$.  When $t> \gamma_2^{-1}$, the values of $p(x;t)$ for all $x<0$ or all $x>0$ will be highly correlated, and relative equilibrium within (but not between) these regions will have been achieved.  Therefore, in this temporal regime $p(x;t)$ can be well-approximated by an expansion within the rank-$2$ (in general, rank-$m$)  \emph{macrostate subspace} spanned by $\varphi_0$ and $\varphi_1$, and only the first two terms in the summation in Eq.\ \eqref{eigenfunction_expansion} need be kept. To obtain a probabilistic description, we replace this truncated eigenfunction expansion by an expansion in the alternative basis provided by the non-negative \emph{macrostate distributions} $\vartheta_a(x)$ and $\vartheta_b(x)$ (to be defined precisely below) shown in Fig.\ \ref{fig:heuristic}A, panels c and d.   $\vartheta_a$ (or $\vartheta_b$) is approximately proportional to  $\varphi_0$ for $x<0$ (or $x>0$) and is approximately 0 for $x>0$ (or $x<0$).  Thus, Eq.\ \eqref{eigenfunction_expansion} can be replaced by the \emph{macrostate expansion}
\begin{equation}
\label{macrostate_expansion}
p(x;t) \approx \sum_\alpha p_\alpha(t) \, \vartheta_\alpha(x) \;,
\end{equation}
where Greek letters index macrostates and sums over Greek letters indicate sums over all macrostates.  (We assume the normalization $\int \vartheta_\alpha(x) \, dx = 1$.)   Since $\vartheta_a$ and $\vartheta_b$ have significant support only for $x<0$ and $x>0$, respectively, $p_a(t)$ and $p_b(t)$ specify the time-dependent amounts of probability in these regions.
Their dynamics are described by the coarse-grained \emph{macrostate master equation}
\begin{equation}
\label{macrostate_master_equation}
\frac{dp_\alpha(t)}{dt} = \sum_\beta p_\beta(t) \, \Gamma_{\beta \alpha} \;,
\end{equation}
where $\Gamma_{\beta \alpha}$ is the \emph{macrostate transition matrix}. As $t \to \infty$, Eq.\ \eqref{macrostate_expansion} reduces to
\begin{equation}
\label{equilibrium_expansion}
\lim_{t\to \infty} p(x;t) = \varphi_0 = \sum_\alpha p^{\rm eq}_\alpha \, \vartheta_\alpha (x) \;.
\end{equation}
where $p^{\rm eq}_\alpha$ is the total probability contained in the macrostate region $\alpha$ at equilibrium.

The $\vartheta_\alpha$ implicitly define the macrostate regions.  To make this explicit, we define \emph{macrostate window functions}
\begin{equation}
\label{continuous_w}
w_\alpha(x) \equiv\frac{ p^{\rm eq}_\alpha \, \vartheta_\alpha(x)}{\varphi_0(x)} \;.
\end{equation}
Eq.\ \eqref{equilibrium_expansion} and the non-negativity of the $\vartheta_\alpha$ imply that
\begin{subequations}
\label{continuous_w_constraints}
\begin{eqnarray}
w_\alpha(x) & \ge 0 \;, \quad \forall \alpha, x \\
\sum_\alpha w_\alpha(x) & = 1 \;, \quad \forall x \;.
\label{continuous_w_summation}
\end{eqnarray}
\end{subequations}
$w_\alpha(x)$ specifies the probability of assignment of microstate $x$ to macrostate $\alpha$. The window functions corresponding to $\vartheta_a$ and $\vartheta_b$ are shown in Fig.\ \ref{fig:heuristic}A.   They define a fuzzy dissection of configuration space into the macrostate regions $x<0$ and $x>0$.

Now we can address the precise definition of the $\vartheta_\alpha$ themselves.  Because they span the macrostate subspace, they must be linear combinations of the slow eigenfunctions:
\begin{equation}
\label{continuous_similarity_transformation}
\vartheta_\alpha(x) = \sum_{n=0}^{m-1} M_{\alpha n} \, \varphi_n(x) \;.
\end{equation}
Since the $\varphi_n$ are smooth, the $\vartheta_\alpha$, and hence the $w_\alpha$, must also be smooth.  This induces an unavoidable uncertainty in microstate assignment.  For example, in Fig.\ \ref{fig:heuristic} the assignments are almost certain for $|x|\gg 0$ where $w_\alpha \approx 1$, but are highly uncertain for $x \approx 0$ where $w_\alpha(x) \approx 0.5$.
The essential point is to choose $M$, and hence the $\vartheta_\alpha$ and $w_\alpha$, so as to maximize  certainty: We define $\Upsilon_\alpha$, the \emph{uncertainty} of macrostate $\alpha$,  as the sum of its equilibrium probability-weighted overlaps with other the other macrostates, relative to its total probability:\footnote{This definition is motivated by analysis of the experimental macrostate preparation and measurement process \cite{Shalloway:96}.}
\begin{equation}
\Upsilon_\alpha  \equiv  \frac{\sum_{\beta \ne \alpha} \int w_\alpha(x) w_\beta(x) p^{\rm eq}(x) \, dx}{\int w_\alpha(x) p^{\rm eq}(x) \, dx}  \;.
\end{equation}
Using Eqs.\ \eqref{equilibrium_expansion}, \eqref{continuous_w}, and \eqref{continuous_w_summation}, the \emph{macrostate certainty} ${\overline \Upsilon}_\alpha$ is
\begin{equation}
\label{bar_Upsilon}
{\overline  \Upsilon}_\alpha \equiv  1 - \Upsilon_\alpha = (p^{\rm eq}_\alpha)^{-1} \, \int w^2_\alpha(x) p^{\rm eq}(x) \, dx  \;.
\end{equation}
We choose $M$ so as to maximize the geometric mean of the ${\overline \Upsilon}_\alpha$ subject to the constraints of Eq.\ \eqref{continuous_w_constraints}. This \emph{minimum uncertainty criterion} minimizes macrostate overlap and, in the example of Fig.\ \ref{fig:heuristic}A, results in the $\vartheta_\alpha$ and $w_\alpha$ shown in panels c and d.  The amount of overlap of these optimized macrostate functions depends on the magnitude of the spectral gap.

\subsection{Adapting macrostate coarse-graining to data clustering}

To adapt the physical coarse-graining procedure to data clustering, we make the computational analogy  \{microstates, macrostates, $\Gamma$\} $\leftrightarrow$ \{items, clusters, $f(D^{-1})$\}. In this analogy, the continuous configuration space of microstates $x$ is replaced by a discrete space of items $i: 1 \le i \le N$, and the probability distribution $p(x,t)$ is replaced by $\bfp(t)$, the  vector of individual item probabilities $p_i(t)$ (e.g., see the simple example in Fig.\ \ref{fig:heuristic}B).  Since $\bfp(t)$ is a probability vector, it must satisfy
\begin{subequations}
\begin{eqnarray}
\label{probability_positivity}
p_i(t) & \ge & 0\;, \quad \forall i,t, \\
\label{probability_normalizataion}
\bfl \cdot \bfp(t) & = & 1\;, \quad \forall t,
\end{eqnarray}
\end{subequations}
where
\[
\bfl_i = 1\;, \quad \forall i \;.
\]

By analogy with Eq.\ \eqref{continuous_time}, we assume that the dynamics in the item-space are described by the microscopic master equation
\begin{equation}
\label{microscopic_dynamics}
\frac{d \bfp(t)}{dt} = \Gamma \cdot \bfp(t)\;,
\end{equation}
where $\Gamma$ is a first-order transition matrix.  Non-negativity of each $p_i(t)$ under time evolution requires that
\begin{equation}
\label{Gamma_off_diagonal}
\Gamma_{ij} \ge 0 \,, \quad i \ne j \;,
\end{equation}
and conservation of probability requires that
\begin{equation}
\label{probability_conservation}
\bfl \cdot \Gamma =0\;.
\end{equation}

The heart of the analogy is to assume that $\Gamma_{ij} \, (i \ne j)$ depends on $D_{ij}$, the dissimilarity between items $i$ and $j$.  If
$D$ were embeddable as a distance matrix in a metric space (e.g., as when it is computed from a measurement matrix $X$), then $\Gamma$ could, in principle, be computed by solving a multidimensional diffusion equation in the continuous space. However, this would be extremely expensive.  Instead, we model $\Gamma$ from $D$ using the following heuristic argument:  A starting point is to set $\Gamma_{ij} = (D_{ij})^{-2}$ by analogy to the rate of diffusion between two isolated microstates in one-dimension.  However, this does not account for the interception of probability flux by intervening items.  To model interception, we use an exponential cutoff scaled to the mean nearest-neighbor squared-distance $\langle d^2_0 \rangle$:
\begin{subequations}
\label{Gamma_D}
\begin{eqnarray}
\Gamma_{ij} & = & \frac{ e^{-(D_{ij})^2/2 \langle d^2_0 \rangle}} {(D_{ij})^2} \;, \quad i \ne j\,, \\
\langle d^2_0 \rangle & = & N^{-1} \sum_{i=1}^N (D_{i <})^2 \;,
\end{eqnarray}
\end{subequations}
where $D_{i<}$ is the smallest element in the $i^{\rm th}$ row of $D$. The diagonal elements of $\Gamma$ are fixed by Eq.\ \eqref{probability_conservation}.

$\Gamma$ defined by Eq.\ \eqref{Gamma_D} is symmetric, so its left- and right-eigenvectors are identical.  Therefore, Eq.\ \eqref{probability_conservation} implies that
\begin{equation}
\label{Gamma_stationary_condition}
\Gamma \cdot \bfl =0 \;,
\end{equation}
and the equilibrium probability vector $\bfp^{\rm eq}$ is
\begin{equation}
\label{equilibrium_vector}
\bfp^{\rm eq} = N^{-1} \, \bfl  \;.
\end{equation}

Eq.\ \eqref{Gamma_off_diagonal} and the symmetry of $\Gamma$ imply that all its eigenvectors $\bfvarphi_n$ are orthogonal and that all its  eigenvalues $-\gamma_n$ are non-positive (see Appendix \ref{appendix:zero_eigenvalues}). It is convenient to use bra-ket notation to indicate the renormalized inner product,
\begin{equation}
\label{inner_product}
\langle \bfx | \bfy \rangle \equiv N^{-1} \, \bfx \cdot {\bfy}   \;,
\end{equation}
and to normalize the  eigenvectors so that
\begin{equation}
\label{normalization}
\langle \bfvarphi_n| \bfvarphi_m \rangle = \delta_{nm} \;.
\end{equation}
Then,
\begin{equation}
\label{phi0_is_1}
\bfvarphi_0 = \bfl\;.
\end{equation}
Fig.\ \ref{fig:heuristic}B illustrates $\bfvarphi_0$ and $\bfvarphi_1$ computed in this way for a simple case of $N=20$ items in a 1-dimensional measurement space.

Because all the elements of $\bfvarphi_0$ are identical, the vector analog of Eq.\ \eqref{continuous_w} is trivial and the macrostate distributions and window functions are directly proportional to each other. Therefore, we simplify by expressing the $m$ \emph{cluster window vectors} directly in terms of the $m$ slow eigenvectors (for now we assume that $m$ has been specified):
\begin{equation}
\label{w_varphi}
\bfw_\alpha = \sum_{n=0}^{m-1} M_{\alpha n} \, \bfvarphi_n \;.
\end{equation}
Analogously to Eqs.\ \eqref{continuous_w_constraints}, the $\bfw_\alpha$ satisfy the positivity and summation constraints required for a probabilistic interpretation:
\begin{subequations}
\label{discrete_window_function_constraints}
\begin{eqnarray}
\label{w_inequalities}
(\bfw_\alpha)_i & \ge&  0 \,, \quad \forall \alpha, i \;, \\
\sum_\alpha \bfw_\alpha & = & \bfl \;.
\label{window_vector_summation}
\end{eqnarray}
\end{subequations}
 Eqs.\ \eqref{phi0_is_1} and \eqref{window_vector_summation} and the orthonormality of the eigenvectors implies the $m$ summation constraints on $M$
\begin{equation}
\label{M_constraints}
\sum_\alpha M_{\alpha n} = \delta_{n0} \;.
\end{equation}

By analogy to Eq.\ \eqref{bar_Upsilon}, the certainty of cluster $\alpha$ is
\begin{equation}
{\overline \Upsilon}_\alpha(M) = \frac{\langle \bfw_\alpha | \bfw_\alpha \rangle}{\langle \bfl | \bfw_\alpha \rangle} \;.
\end{equation}
As in the continuous case, $0 \le {\overline \Upsilon}_\alpha \le 1$. Maximizing the geometric mean of the ${\overline \Upsilon}_\alpha$ is equivalent to minimizing the objective function
\begin{equation}
\label{objective_function}
\Phi(M) \equiv - \sum_\alpha \log {\overline \Upsilon}_\alpha(M) \;.
\end{equation}

Minimization of $\Phi$ consistent with the linear constraints of Eq.\ \eqref{M_constraints} and the linear inequality constraints of Eq.\ \eqref{w_inequalities} fixes $M$, and hence the  $\bfw_\alpha$, for a specified value of $m$.  The solution of this global optimization problem is discussed in Appendix \ref{appendix:global_minimization}.  There we show that the resultant $\bfw_\alpha$ are linearly independent, so they provide a complete basis for the macrostate subspace.  Once the $\bfw_\alpha$ have been computed we complete the clustering procedure for $m$ by assigning each item $i$ to the cluster $\alpha$ having the maximal value of $(w_\alpha)_i$.  We say that the assignment is ``strong'' or ``weak'' depending on how close this maximal value, the \emph{item assignment strength}, is to 1.  The assignments are extremely strong for the example shown in Fig.\ \ref{fig:heuristic}B (re panels c and d) because of the relatively large separation between the two clusters.

In some cases, the procedure may define a cluster with only a single item. In this case $\tau^{\rm local}$ is undefined and there is no meaningful dissection of dynamics into slow and fast processes.  Therefore we treat such outliers by a special procedure: When one is identified, it is removed from the dataset and the pruned dataset is reanalyzed. The pruning procedure is repeated if new outliers appear.  We designate the final clustering as $\mathcal{C}(m)$.

\subsection{Determining the number of clusters}

We use two conditions to determine if $\mathcal{C}(m)$ is an \emph{acceptable clustering}: As motivated above, we examine the eigenspectrum of $\Gamma$ for spectral gaps, which are defined as
\begin{equation}
\label{gap_condition}
\gamma_m/\gamma_{m-1} > \rho_\gamma \;,
\end{equation}
where $\rho_\gamma$ is the \emph{minimum gap parameter}. However, Eq.\ ({\ref{gap_condition}) alone may accept excessively fuzzy clusters having weak item assignment vectors.  To eliminate these, we supplement Eq.\ ({\ref{gap_condition}) with the \emph{minimum macrostate certainty conditions}:
\begin{equation}
\label{minimum_certainty_condition}
{\overline \Upsilon}_\alpha > \rho_\Upsilon \,,\quad \forall \alpha \;.
\end{equation}
We have found that choosing $\rho_\gamma = 3$ and $\rho_\Upsilon= 0.68$ (the fraction of the normal distribution contained within one standard-deviation of the mean) works well for all the problems that we have tested (see Results).

The complete algorithm is to sequentially compute $\mathcal{C}(m)$ for $m=2,\, 3,\, \ldots$ and to test these clusterings for  acceptability according to Eq.\ \eqref{minimum_certainty_condition}.  If multiple clusterings are acceptable, we will usually be most interested in the one of largest $m$, since it provides the finest resolution.  As a practical matter, if $\mathcal{C}(m)$ is not acceptable for three consecutive $m$'s we assume that it will not be acceptable for higher $m$'s and terminate the analysis.

\subsection{Computational implementation}

Only two steps in the procedure are potentially expensive: computing the slow eigenvectors and eigenvalues of $\Gamma$ and the global minimization of $\Phi(M)$. Since we only use a relatively small number (typically $m < 10$) of slow eigenvectors, it suffices to compute only these via the Lanczos method \cite{Golub:89} at cost $\sim O(N^2)$.  The global minimization of $\Phi(M)$ is a linearly constrained global optimization problem in $m(m-1)$ dimensions. The number of vertices of the feasible region-bounding polytope increases with $N$, formally as a polynomial dependent on $m$.  However, at least for the problems tested here, a simple solver is adequate (see Appendix \ref{appendix:global_minimization}).

\section{Results}
We tested  the method on a number of problems that have challenged other clustering methods. Bivariate problems in which the data set can be graphically displayed in two dimensions were used to enable subjective evaluation of the quality of the results.  In addition, to check that performance did not depend on low dimensionality of the data space, we tested problems where the items were embedded in a 20-dimensional space.

\subsection{Bivariate test-cases}

The algorithm was evaluated on four previously described difficult test-cases. In each case, the dataset consisted of $N_M = 2$ measurements on each of $N$ items.  These can be represented as $N$ points in a 2-dimensional space.  For example, the ``crescentric'' clustering problem shown in Fig.\  \ref{fig:crescentric}a consists of 52 items, each represented as a point in the 2-dimensional measurement space.  The two clusters are closely juxtaposed crescents, which makes the problem difficult \cite{Wong:83,Everitt:01}. The $D$ matrix was computed from the coordinates using Eq.\ \eqref{D} with $g_{ab}= \delta_{ab}$, and $\Gamma$ was computed from $D$ according to Eqs.\ \eqref{Gamma_D}. The slowest eigenvectors, $\bfvarphi_0$, $\bfvarphi_1$, and $\bfvarphi_2$, are graphically displayed in panels b, c and d, respectively.  As per Eq.\ \eqref{equilibrium_vector}, all components of $\bfvarphi_0$ are identical.  It is gratifying to see that $\bfvarphi_1$ clearly reflects the two-cluster structure: the components of all the items in the bottom-right crescent are positive, while the components of all the items in the other crescent are negative.  The next eigenvector, $\bfvarphi_2$, has three localized regions of same-sign components. Subjectively, it is clear that separating into these regions would overdissect the space.  As predicted by the discussion above, these eigenvector properties in the spatial domain correspond in the time domain to an eigenspectrum gap between $\gamma_1$ and $\gamma_2$ (Fig.\ \ref{fig:crescentric} and Table \ref{table:crescentric}). In contrast, there is no gap between $\gamma_2$ and $\gamma_3$ (Fig.\ \ref{fig:crescentric}).  This suggests that the $m=2$, but not the $m=3$ clustering will be acceptable.

The task for the algorithm is to recognize that the correct clustering is embedded in the structure of $\bfvarphi_1$, and to define the proper clustering.  Applying it for $m=2, \, 3, \ldots$ yields the clusters shown in the top panels of Fig.\ \ref{fig:2d_examples}.  (For illustration, we display clusterings that do not satisfy the spectral gap condition, even though these would not be computed by an efficient algorithm.)  The  cluster certainties ${\overline \Upsilon}_\alpha$ are listed in Table \ref{table:crescentric}. The $m=2$ clustering satisfies both Eqs.\ \eqref{gap_condition} and \eqref{minimum_certainty_condition}, and all clusterings with $m > 2$ fail both criteria.  Therefore, the algorithm correctly selects $m=2$ clusters.  The individual item assignment strengths for this clustering are displayed in Fig.\ \ref{fig:assignment_strengths}; most are in the range of $0.7-0.9$, indicating that there is significant fuzziness resulting from the close juxtaposition of the clusters. Nonetheless, all the item assignments are made correctly.

Three other test problems were analyzed in the same way: (1) The ``intersecting'' problem consists of two barely-contacting sets of items having highly anisotropic Gaussian distributions.  It has previously been used to demonstrate the weakness of
non-parametric optimization clustering for clusters of greatly different shapes and sizes \cite{Everitt:01}. (2) The ``parallel'' problem consists of two highly extended, anisotropic sets of items whose separation along the vertical axis is much smaller than their horizontal extent. It has previously been used to demonstrate the failure of agglomerative hierarchical methods \cite{Everitt:01}.  (3) The ``horseshoe'' problem \cite{Szallasi:01} consists of a central cluster of items surrounded by a horseshoe-shaped cluster of items.  The center-of-mass of the outer cluster lies within the inner cluster, increasing difficulty. In addition, a ``random'' test set, in which points were randomly distributed within a square two-dimensional region, was analyzed as a control.

The results obtained for $m=2$, 3, 4, and 5 are illustrated in Fig.\ \ref{fig:2d_examples}. The acceptable clusterings that  satisfy Eqs.\ \eqref{gap_condition} and \eqref{minimum_certainty_condition} are outlined by dark boxes.  Only a single clustering is acceptable in each case (although this need not be so in general).  None of the random control clusterings are acceptable, correctly indicating that it should not be clustered.

As with the crescentric problem, the clustering solution for the ``horseshoe'' test-case (fourth row, Fig.\ \ref{fig:2d_examples}) is straightforward, with $m=2$.  Cluster certainties  (Table \ref{table:2d_examples}) and item assignment strengths (Fig.\ \ref {fig:assignment_strengths}) are extremely strong ($> 0.99$).  The ``parallel'' problem is slightly more challenging in that two of the items (located at the extreme left and right sides of the item distributions) are identified as outliers.  Nonetheless, the algorithm correctly identifies the $m=2$ clustering of the non-outlying items. As expected, the item assignment strengths are lower for the items in the central overlapping region, and higher for the non-overlapping items near the left and right edges (Fig.\ \ref{fig:assignment_strengths}).

The solution to the ``intersecting'' problem is more elaborate:  While the $m=2$ solution is subjectively acceptable, the assignment strengths of some of the items in the vicinity of the intersection have weak item assignment strengths.  Because of this, the $m=2$ and $m=3$ clusterings do not satisfy the required assignment certainty condition Eq.\ \eqref{minimum_certainty_condition} and are rejected by the algorithm. The acceptable $m=4$ clustering resolves this difficulty by segregating these uncertain items into a separate small cluster.  It also segregates two outliers (in the top-right corner) while assigning most of the items to two major clusters, as desired.  The individual item assignment strengths are strong, except for one item near the intersection of the three clusters (Fig.\ \ref{fig:assignment_strengths}).

None of the $\mathcal{C}(m)$ are acceptable for the ``random'' distribution of items because all of the $\gamma_m/\gamma_{m-1}$ were $<2.5$ for $m>1$.  Thus, the algorithm is not fooled into spurious clustering.

\subsection{Gaussians with varying overlap in two and twenty dimensions}

We systematically tested the performance of the algorithm as a function of the relative distance between clusters. For this purpose, four pseudo-random groups of 50 items were generated with Gaussian kernels having variance $\lambda^2_\ell$. The centers-of-mass of the groups were themselves pseudo-randomly selected from a Gaussian kernel having variance $\lambda^2_g$ (see Fig.\ \ref{fig:2d_gaussian}). The corresponding ratio of the expected root-mean-square (rms) \emph{inter}cluster item-item separations to the rms \emph{intra}cluster item separations is
\begin{equation}
\label{inter_intra}
\sqrt{\frac{\langle (\Delta R)^2 \rangle_{\rm inter}}{\langle (\Delta R)^2 \rangle_{\rm intra}}} = \sqrt{(\lambda^2_\ell + \lambda^2_g)/\lambda^2_\ell} \;,
\end{equation}
Tests in a bivariate measurement space were conducted for $\lambda_g/\lambda_\ell$ varying from 16 (where the clusters were highly separated) down to 2 (where the clusters were completely overlapping).  The algorithm dissects the items into four clusters when  $\lambda_g/\lambda_\ell=16$.  When $\lambda_g/\lambda_\ell=8$ and $\lambda_g/\lambda_\ell=4$, the top two groups partially merge (see Fig.\ \ref{fig:2d_gaussian}), and the algorithm accordingly reports $m=3$ clusters.  The clusters are not subjectively separable for $\lambda_g/\lambda_\ell=2$; correspondingly, the algorithm reports $m=1$ cluster. The assignment strengths for these clusterings are displayed in Fig.\ \ref{fig:2d_gaussian}.

The same test was performed with four groups generated as described above using Gaussian kernels in a 20-dimensional space.  The increased dimensionality does not alter Eq.\ \eqref{inter_intra}. However, the distributions of the inter- and intra-group
squared-distances are narrower: the standard deviations of the inter- and intra-group $(\Delta R)^2$ normalized by their means are both smaller by factors of $\sqrt{20/2} = \sqrt{10}$.  Therefore, for a given value of $\lambda_g/\lambda_\ell$, clustering is actually easier in higher dimensionality.  To compensate and make the 20-dimensional test more challenging, the range of $\lambda_g/\lambda_\ell$ was reduced by a factor of 4 (roughly matching $\sqrt{10}$); i.e.,  $\lambda_g/\lambda_\ell$ was varied from 4 down to 0.5.  The algorithm correctly identifies the four clusters for $\lambda_g/\lambda_\ell = 4$ and $\lambda_g/\lambda_\ell = 2$.  The individual item assignment strengths of these clusterings are displayed in Fig.\ \ref{fig:20d_gaussian}.  These are all close to one for $\lambda_g/\lambda_\ell = 4$ and $\lambda_g/\lambda_\ell = 2$, indicating unambiguous clustering.  At smaller values of $\lambda_g/\lambda_\ell$, the only clustering satisfying both the minimum gap and minimum certainty conditions has one cluster containing all the items. Even so, for $\lambda_g/\lambda_\ell=1$, the (formally unacceptable) $m=3$ clustering correctly reflects some of the group structure (Fig.\ \ref{fig:20d_gaussian}).

\section{Discussion}

We have shown that macrostate clustering performs well on a variety of test problems that have challenged other methods.  The method only needs a dissimilarity matrix $D$ (not a data matrix $X$) and has the advantage of being non-hierarchical,\footnote{For example, the $m=5$ ``crescentric'' clustering can not be obtained by subdividing its $m=4$ clustering and the $m=4$ ``horseshoe'' clustering is not hierarchically related to its $m=3$ clustering.  Nevertheless, inherent hierarchical structure can still emerge, and some was evident in all the problems.  For example, all the clusterings for $2 \le m \le 5$ for the ``intersecting'' and ``parallel'' problems are hierarchically related (Fig.\ \ref{fig:2d_examples}).} which should improve performance in general.  Beyond identifying potential clusterings, it uses internal criteria---the eigenspectrum gaps $\gamma_m/\gamma_{m-1}$ and the cluster certainties $\overline{\Upsilon}_\alpha$---to determine the appropriate number of clusters.   The corresponding acceptance parameters, $\rho_\gamma$ and  $\rho_\Upsilon$, were empirically determined and gave robust performance---a single choice worked well for all the problems tested.

Eigenvectors have previously been used for clustering by many different spectral graph theory (SGT) partitioning methods:  SGT bipartitioning methods use the values of $\bfvarphi_1$ to define a one-dimensional ordering of the items which can then be divided by a heuristic. A variety of different approaches have been developed to extend this to multiple eigenvectors and clusters (\cite{Seary:95,Spielman:96,Alpert:99,Weiss:99}, for review).  For example, recursive spectral bipartitioning generates a hierarchical binary tree \cite{Barnard:94}; some methods use $k$ eigenvectors to define $2^k$ clusters  \cite{Hendrickson:93}; and many methods project the items into the subspace spanned by $k$ eigenvectors and then use a partitioning heuristic to identify clusters within the subspace (e.g., \cite {Pothen:90,Alpert:99,Weiss:99,Kannan:01,Meila:01,Ng:01} and references therein).

Macrostate clustering differs in that it computes continuous (fuzzy) assignment window vectors rather than partitionings.\footnote{Drineas et al.\ \cite{Drineas:99} consider real-valued ``generalized clusters'' within a SGT context, but
these are indefinite and do not have a probabilistic interpretation.}  This has important ramifications: It permits the
window vectors to be expressed as linear combinations of the eigenvectors [see Eq.\ \eqref{w_varphi}]; this necessarily results
in window function overlap and cluster uncertainty. Combining these concepts with the principle of uncertainty minimization
provides a simple prescription for the concurrent use of multiple eigenvectors in clustering. A related difference is that the
number of clusters is internally determined in macrostate clustering, while it is usually fixed \emph{a priori} or
determined by eigensystem-independent heuristics in SGT methods (e.g., \cite{Seary:95,Alpert:99,Weiss:99} and references therein).
It is perhaps surprising that the spectral gap condition has not been used for this purpose in SGT approaches.\footnote{However,
spectral gaps have been used heuristically to determine the appropriate dimensionality of singular subspaces in data mining
\cite{Azar:01}.}  This may reflect the fact that it does not work well by itself, and the companion minimum cluster certainty
condition is not available when (crisply) partitioning. Macrostate and SGT clustering also differ in the manner in which
$\Gamma$ (or the SGT analog) is computed from the dissimilarity matrix $D$.  SGT methods typically use a weight matrix equivalent
to $\Gamma_{ij} = \exp(-D_{ij}/\sigma)$, $i \ne j$, where $\sigma$ is an empirically-determined scale constant. In contrast,
motivated by the analogy to a diffusive system, we used Eqs.\ \eqref{Gamma_D}.  While this difference not of fundamental
significance, the relationship between $\Gamma$ and $D$ can affect performance.  Thus, it may be helpful to test the use of Eqs.\
\eqref{Gamma_D} in SGT methods or the SGT relationship in macrostate clustering.

The use of a linear transformation from indefinite, orthogonal eigenvectors to semidefinite, non-orthogonal window  vectors is fundamental, but some freedom remains in the choice of the objective function used to determine the optimal transformation and in the conditions used to determine acceptable clusterings.  An uncertainty minimization criterion is a natural choice, since it is (in an information-theoretic sense) the entropic counterpart to the (implicit) ``energy'' minimization criterion that focuses attention on the slow eigenvectors (see Sec.\ II.C of \cite{Ulitsky:98}).  On-the-other-hand, the definition of uncertainty could be modified and tested for improved performance.  Similarly, while we believe that it is advantageous to combine energetic (spectral gap) and entropic (cluster certainty) conditions in determining the number of clusters, it may be possible to improve upon the specific criteria used here.

Other improvements and extensions merit attention: (1) While we accepted or rejected each clustering \emph{in toto}, it may be useful in some cases to examine incomplete clusterings in which only some of the clusters satisfy the cluster certainty condition. This modification would enable the algorithm to resolve all four clusters for the case of $\lambda_g/\lambda_\ell=8$ in Fig.\ \ref{fig:2d_gaussian}.\footnote{The $m=5$ solution identifies the four major clusters with strong certainty, but also groups three items (located near the boundary between the two top clusters) into a fifth cluster which has $\overline{\Upsilon}_\alpha < \rho_\Upsilon$.  In an incomplete clustering, all but these three items would be unambiguously assigned.} (2) The individual item assignment strengths $(\bfw_\alpha)_i$ measure the certainty of each assignment, but their precise statistical significance is not known.  It would be helpful to have a model for assessing this.  (3) The \emph{cluster transition matrix} $\gamma_{\beta \alpha}=
\langle \bfw_\beta | \Gamma | \bfw_\alpha \rangle$ can be used to assess the strength of the relationship between the clusters and may be useful in setting the cluster acceptance criteria. 
(4) We have defined $\Gamma$ as a symmetric matrix, which implies that  $\bfp^{\rm eq} \propto \bfl$. However, this restriction is not required: The generalization to asymmetric $\Gamma$ is straightforward \cite{Shalloway:96} and it could be used to incorporate additional experimental information.  For example, if item $i$ is known \emph{a priori} to be partially redundant with other items (e.g., when analyzing expression levels of members of gene families), it may be given reduced weight in the analysis by setting  $\bfp^{\rm eq}_i < 1$.

Our main goal has been a proof-of-principle demonstration of the high quality of the clusterings provided by the dynamical macrostate approach. The current implementation is sufficiently efficient for problems where $N \sim O(10^2)$, but we have not examined performance for very large problems.  The continuous formulation replaces the NP-hard combinatoric SGT partitioning problem with a global minimization problem having polynomial complexity in $N$.  However, the order of the polynomial can be very large for large $m$ (Appendix \ref{appendix:global_minimization}) so, formally, this is not much of an improvement.  Nonetheless, as discussed in Appendix \ref{appendix:global_minimization}, because the objective function is smooth and the constraints are highly degenerate, a simple solver has worked well and we believe that it will be possibly to obtain adequate approximate solutions efficiently even for very large problems. This remains to be examined.

\begin{acknowledgments}
We thank Bruce Church, Jason Gans, Ron Elber, Jon Kleinberg, and Golan Yona for helpful conversations and the NSF (grant CCR9988519) and the NIH (training grant T32GM08267) for support.
\end{acknowledgments}

\appendix

\section{Minimizing $\Phi(M)$}
\label{appendix:global_minimization}

$\Phi(M)$ is to be minimized as a function of the $m^2$ elements of $M_{\alpha n}$ within the feasible region specified by the $m \times N$ linear inequality constraints of Eq.\ \eqref{w_inequalities}.  The rows of $M$ can be regarded as the coordinates of $m$ particles in the $m$-dimensional space of the slow eigenvectors. Designating the coordinate row vector of particle $\alpha$ as $\overrightarrow{M}_\alpha = (M_{\alpha 0}, \,M_{\alpha 1},\, \ldots \, M_{\alpha (m-1)})$, $M$ is the outer product of the $\overrightarrow{M}_\alpha$'s:
\begin{equation}
\label{outer_product}
M= \bigotimes_\alpha \overrightarrow{M}_\alpha\;.
\end{equation}
The equality constraints of  Eq.\ \eqref{M_constraints} imply that the center-of-mass of the $m$ particles is at position
\begin{equation}
\label{generalized_M_equation}
\frac{1}{m} \sum_\alpha \overrightarrow{M}_\alpha = \frac{\hat{\varepsilon}_0}{m} \;,
\end{equation}
where $\hat{\varepsilon}_0$ is the unit vector in the $0^{\rm th}$ direction:
\begin{equation}
\label{epsilon_0} \hat{\varepsilon}_0 = (1, \, 0,\, \ldots, 0)\;.
\end{equation}
[Eq.\ \eqref{epsilon_0} must be modified when there is more that one stationary eigenvector; see Appendix \ref{appendix:zero_eigenvalues}.] The feasible region is a polytope in the $m(m-1)$ dimensional subspace where Eq.\ \eqref{generalized_M_equation} is satisfied.

The minimum of $\Phi(M)$ must lie at a vertex of this polytope. \emph{Proof}: The gradient of $\Phi$ with respect to $\overrightarrow{M}_\alpha$ is
\begin{equation}
\label{gradient}
\overrightarrow{\nabla}_\alpha \Phi \equiv \frac{\delta \Phi}{\delta \overrightarrow{M}_\alpha} =
-2 \frac{\overrightarrow{M}_\alpha}{|\overrightarrow{M}_\alpha|^2} + \frac{\hat{\varepsilon}_0}{\overrightarrow{M}_\alpha \circ \hat{\varepsilon}_0} \;,
\end{equation}
and the Hessian is
\begin{equation}
\label{Hessian}
\overrightarrow{\nabla}_\alpha \otimes \overrightarrow{\nabla}_\beta \Phi \equiv
\frac{\delta^2 \Phi}{\delta \overrightarrow{M}_\alpha \, \delta \overrightarrow{M}_\beta} =
- \delta_{\alpha \beta} \left[
\frac{2 I}{|\overrightarrow{M}_\alpha|^2}
- 4 \frac{\overrightarrow{M}_\alpha \otimes \overrightarrow{M}_\alpha}{|\overrightarrow{M}_\alpha|^4}
+ \frac{\hat{\varepsilon}_0 \otimes \hat{\varepsilon}_0} {(\overrightarrow{M}_\alpha \circ \hat{\varepsilon}_0)^2} \right] \;,
\end{equation}
where $I$ is the $m \times m$ identity matrix and $\circ$ denotes the inner product over the eigenvector indices,
\[
\vec{x} \circ \vec{y} \equiv \sum_{n=0}^{m-1} x_n \, y_n \;.
\]
The gradient does not vanish anywhere, so $\Phi$ has no minimum in the absence of constraints.

In fact, a minimum will occur only when \emph{all} $m^2$ degrees-of-freedom are constrained by the $m$ equality constraints and $m(m-1)$ inequality constraints.  To see this, consider the situation without the equality constraints, but with some number $c \le m(m-1)$ of active inequality constraints. Each active inequality constraint acts on a single $\bfw_\alpha$, so by Eq.\ \eqref{w_varphi} it acts on a single $\overrightarrow{M}_\alpha$ to enforce
\begin{equation}
\label{constrained_outer_product}
\overrightarrow{M}_\alpha \circ  (\overrightarrow{\bfvarphi})_i = 0 \;,
\end{equation}
where $\overrightarrow{\bfvarphi}$ is the supervector having components $(\bfvarphi_0, \; \bfvarphi_1,\, \ldots,\; \bfvarphi_{m-1})$. Therefore, the inequality constraints are separable and, similarly to Eq.\ \eqref{outer_product}, the space of inequality-constrained $M$'s can be expressed as the outer-product of the subspaces of inequality-constrained $\overrightarrow {M}_\alpha$'s. Thus, if $M^c=\bigotimes_\alpha M^c_\alpha$ is an inequality-constrained  minimizer of $\Phi$, it must be stable with respect to independent variations of each of the inequality-constrained $M^c_\alpha$.  However, this is not possible:  For any such variation $\overrightarrow{M}^c_\alpha \to \overrightarrow{M}^c_\alpha +\vec{\delta}_\alpha$, the existence of a minimum would require that
\begin{equation}
\label{grad_condition}
\vec{\delta}_\alpha \circ \overrightarrow{\nabla}_\alpha \Phi = 0
\end{equation}
and
\begin{equation}
\label{hessian_condition}
\vec{\delta}_\alpha \circ (\overrightarrow{\nabla}_\alpha \otimes \overrightarrow{\nabla}_\alpha) \Phi \circ \vec{\delta}_\alpha > 0 \;.
\end{equation}
However, Eqs.\ \eqref{gradient} and \eqref{grad_condition} imply that
\[
\frac{\overrightarrow{M}_\alpha \circ \vec{\delta}_\alpha}{|\overrightarrow{M}_\alpha|^2} = \frac{\vec{\delta}_\alpha \circ{\hat{\varepsilon}_0}}{2 \, \overrightarrow{M}_\alpha \circ \hat{\varepsilon}_0} \;,
\]
and combining this with Eq.\ \eqref{Hessian} implies that
\[
\vec{\delta}_\alpha \circ (\overrightarrow{\nabla}_\alpha \otimes \overrightarrow{\nabla}_\alpha) \Phi \circ \vec{\delta}_\alpha =
- \frac{2 |\vec{\delta}_\alpha|^2}{|\overrightarrow{M}_\alpha|^2} <0 \;.
\]
Thus, Eqs.\ \eqref{grad_condition} and \eqref{hessian_condition} cannot both be true. Therefore, a minimum can occur only if \emph{all} variations of the $M_\alpha$ are prevented by   a combination of inequality- and equality-constraints.  Since there are only $m$ equality constraints, we must have $c=m(m-1)$ active inequality constraints. This corresponds to a vertex of the feasible region.

Note also that the minimizing $\{\overrightarrow{M}^c_\alpha\}$ must be linearly independent within the $m$-dimensional slow eigenvector space.  This implies that the associated $\{\bfw_\alpha\}$ must span the macrostate subspace. \emph{Proof}:  If the $\{\overrightarrow{M}^c_\alpha\}$ are not independent, there would exist a linear combination of vectors such that
\[
\sum_{\alpha} \xi_\alpha \, \overrightarrow{M}^c_\alpha = 0 \;.
\]
Then, the combined variation
\[
\overrightarrow{M}^c_\alpha \to \overrightarrow{M}^c_\alpha + \delta \, \xi_\alpha \overrightarrow{M}^c_\alpha \;, \qquad \forall \alpha,
\]
where $\delta$ is a small number, will not affect the equality constraint Eq.\ \eqref{generalized_M_equation}. As proven above, all the components of the constrained minimum must be fixed by constraints, so this variation must be excluded by an inequality constraint.   However, this variation only rescales each $\overrightarrow{M}^c_\alpha$ and hence each $\bfw_\alpha$. Therefore it also will not affect the inequality constraints and is permitted, contrary to assumption. \emph{Reductio ad absurdum.}

To find the vertex with the lowest value of $\Phi$, we used a simple minimizer that operates in the $m(m-1)$-dimensional
subspace that remains after one of the $M_\alpha$ has been explicitly eliminated using  Eq.\ \eqref{generalized_M_equation}.
The minimizer starts from $\overrightarrow{M}_\alpha= \hat{\varepsilon}_0 /m$, $\forall \alpha$, chooses a random
direction in the $m(m-1)$-dimensional space, proceeds to the nearest inequality constraint, and then proceeds along faces of
the feasible region (of decreasing dimensionality) until a vertex is reached. This process was repeated until the same extremal
minima was found three times or for a minimum of 500,000 trials, whichever was greater.

Accounting for the separability of the inequality constraints and assuming no degeneracies between the values of the $\bfvarphi_n$ (the usual case), the number of vertices of the constraining polytope might grow as rapidly as $O(N^m)$.  However, we expect that most of the inequality constraints of Eq.\ \eqref{w_inequalities} will be almost degenerate because of the relatively small differences between the components of the eigenvectors at different items within a cluster.  Moreover, the objective function $\Phi$ is smooth, so we expect that the variation of $\Phi$ over nearby vertices will be small.  Therefore, it will not much affect the $\bfw_\alpha$ if a neighbor, rather than the global minimizer itself, is found. Thus, we anticipate that the practical growth in computational cost with $N$ will be much less than the worst-case bound. These considerations also suggest that it will always be advantageous to use solvers that move through the $[m(m-1)-1]$-dimensional space of search-space directions rather than between vertices of the constraining polytope.

\section{Degenerate ``zero'' eigenvalues}
\label{appendix:zero_eigenvalues}

Because $\Gamma$ is a symmetric matrix which satisfies Eqs.\ \eqref{probability_conservation} and \eqref{Gamma_stationary_condition},
\begin{equation}
\label{Gamma_isomorphism}
- {\bfx} \cdot \Gamma \cdot {\bfx}=
 \sum_{\stackrel{j>i}{i}} \Gamma_{ij} \, (x_i-x_j)^2 \;,
\end{equation}
for any vector $\bfx$. The right-hand-side (rhs) can be viewed as the potential energy of $N$ particles having pairwise quadratic interactions in one-dimension.  Because all the off-diagonal elements of $\Gamma$ are positive, the rhs must be non-negative.  The implied non-positivity of ${\bfx} \cdot \Gamma \cdot {\bfx}$ for all $\bfx$ implies that all the eigenvalues of $\Gamma$ must be non-positive. Furthermore, the isomorphism makes it evident that $\bfx = \bfl$ is the only stationary eigenvector (up to a multiplicative constant) unless the dataset contains an \emph{isolated subset} $\mathcal{S}$, which has $\Gamma_{ij}=0$ if $i \in \mathcal{S}$ and $j \not \in \mathcal{S}$.  In this case, $\Gamma$ will have multiple zero eigenvalues, and there will be one stationary eigenvector corresponding to each isolated subset. This degeneracy can be removed by analyzing each isolated subset independently.

It is more common to encounter approximate isolation in which none of the $\Gamma_{ij}$ is exactly zero but in which there are multiple small eigenvalues that are $~0$ on the scale of numerical accuracy.  (This occurs in the Gaussian clustering problem shown in Fig.\ \ref{fig:2d_gaussian} when $\lambda_g/\lambda_\ell$ is large.)  This can cause numerical problems: the $\bfvarphi_0$ returned by a numerical eigensystem solver will not necessarily satisfy Eq.\ \eqref{phi0_is_1}, but instead will be a linear combination of the approximately degenerate eigenvectors.  Because of this, Eq.\ \eqref{phi0_is_1}, and hence Eq.\ \eqref{M_constraints}, may not be true.

The simplest resolution of this numerical problem is to replace Eq.\ \eqref{M_constraints} with
Eq.\ \eqref{generalized_M_equation} and to replace Eq.\ \eqref{epsilon_0} with
\begin{equation}
\label{modified_epsilon_0} \hat{\varepsilon}_0 = \langle
\bfl |\overrightarrow{\bfvarphi} \rangle \;.
\end{equation}
This does not require the numerical validity of Eq.\ \eqref{phi0_is_1}.


\newpage

\begin{figure}[h]
\caption{\label{fig:heuristic}
Heuristic examples. (A) Identifying the macrostates of a continuous stochastic system in one-dimension. a: The potential $V(x)$ and eigenvalue spectrum. b: The zeroth and first excited right-eigenfunctions of the corresponding diffusive dynamical (Smoluchowski) equation. c and d: The two macrostate distribution and window functions. (B) Macrostate clustering of items in a one-dimensional space. a: The positions of the items in the univariate measurement space. b: Graphical representation of the zeroth and first eigenvectors of $\Gamma$; the height of the bar at the position of item $i$ corresponds to its component within the indicated eigenfunction. c and d: The components of the two window vectors corresponding to the left ($\bfw_a$) and right ($\bfw_b$) clusters.}
\end{figure}

\begin{figure}[h]
\caption{\label{fig:crescentric}
``Crescentric'' clustering problem and its slow eigenvectors. a: The $x$ and $y$ coordinates of each point correspond to two measurement values of the corresponding item. b, c and d: $\bfvarphi_0$, $\bfvarphi_1$, and $\bfvarphi_2$, respectively. For illustration, the amplitude of the $i^{\rm th}$ component of each $\bfvarphi_n$ is represented by the height (if positive) or depth (if negative) of a cone centered at position $i$. The relative magnitudes of the corresponding eigenvalues are indicated.}
\end{figure}

\begin{figure}[h]
\caption{\label{fig:2d_examples}
Bivariate test-cases. The algorithmically-determined clusterings $\mathcal{C}(m)$ for $2 \le m \le 5$ are displayed for four bivariate examples in which the items are points in a two-dimensional measurement space.  Clusters are distinguished by different symbols, except that unfilled squares identify items that were designated as outliers by the algorithm.  The acceptable clusterings, which satisfy Eqs.\ \eqref{gap_condition} and \eqref{minimum_certainty_condition}, are outlined by dark boxes.}
\end{figure}

\begin{figure}[h]
\caption{\label{fig:assignment_strengths}
Item assignment strengths for the acceptable clusterings.  The acceptable clusterings for each of the problems in Fig.\ \ref{fig:2d_examples} are shown.  The height of the dark section of the bar relative to its total height at the position of an item indicates its assignment strength.}
\end{figure}

\begin{figure}[h]
\caption{\label{fig:2d_gaussian}
Clustering of Gaussian-distributed items in two dimensions for various cluster separations.  Top: The unique acceptable clustering for each value of $\lambda_g/\lambda_\ell$ is indicated. Bottom: The height of the dark section of the bar at the position of an item indicates its assignment strength. (Most of the strengths are $\approx 1$).}
\end{figure}

\begin{figure}[h]
\caption{\label{fig:20d_gaussian}
Item assignment strengths for cluster solutions for various group separations in 20 dimensions. Items were pseudo-randomly distributed into four groups in a 20-dimensional measurement space for different values of $\lambda_g/\lambda_\ell$ as described in the text.  The items within each group have consecutive serial numbers (i.e., items 1--50 are in the first group, 51--100 are in the second group, etc.). Their assignment strengths for the indicated $\mathcal{C}(m)$ clusterings are displayed in the each case. (Item 171 is an outlier for both clusterings shown in the bottom row; hence it is not assigned to any cluster.)  However, only the $m=4$ clusterings for $\lambda_g/\lambda_\ell= 4$ and $\lambda_g/\lambda_\ell=2$ are acceptable; the $\mathcal{C}(3)$ and $\mathcal{C}(2)$ shown in the bottom panels fail the acceptability conditions of Eqs.\ \eqref{gap_condition} and \eqref{minimum_certainty_condition} because of their low cluster certainties.}
\end{figure}

\begin{table}[h]
\caption{\label{table:crescentric}
Crescentric cluster analysis}
\begin{center}
\begin{tabular}{ccc}
\multicolumn{1}{c}{m} &
\multicolumn{1}{c}{\;\;$\frac{\gamma_{m}}{\gamma_{m-1}}$\;\;} &
\multicolumn{1}{c}{$\overline{\Upsilon}_\alpha(m)$} \\
\hline  
{\bf 2}       & 3.52                        & 0.71 \\
 & & 0.70 \\
{\bf 3}       & 1.12                        & 0.67 \\
 & &                                 0.41 \\
 & &                                 0.53 \\
{\bf 4}       & 2.73                        & 0.83 \\
 & &                                 0.81 \\
 & &                                0.51 \\
 & &                                0.53 \\
{\bf 5}       & 1.03                        & 0.71 \\
 & &                                 0.47 \\
 & &                                0.55 \\
 & &                                0.38 \\
 & &                                0.38 \\
\end{tabular}
\end{center}
\newpage
\end{table}

\begin{table}[h]
\caption{\label{table:2d_examples}
Bivariate test case analyses}
\begin{center}
\begin{tabular}{lcrl}
\multicolumn{1}{c}{Problem} &
\;\;\;  m \;\;\;
&
\multicolumn{1}{c}{$\frac{\gamma_{m}}{\gamma_{m-1}}$} &
\multicolumn{1}{c}{\;\;$\overline{\Upsilon}_\alpha(m)$} \\
\hline
{\bf Crescentric} & 2 & 3.52 & \;\;0.71 \\
 & && \;\;0.70 \\
{\bf Intersecting} & 4           & 3.82                        & \;\;0.91
\\
 & &&                                        \;\;0.95 \\
 & &&                                        \;\;0.84 \\
 & &&                                        \;\;0.94 \\
{\bf Parallel}   & 2         & 10.68                       & \;\;0.93 \\
 & &&                                        \;\;0.93 \\
{\bf Horseshoe}   &2        & 60.73                       & \;\;0.998 \\
 & && \;\;0.99 \\
{\bf Random}  & 1           & --\;\;                        & \;\;\;\;-- \\
\end{tabular}
\end{center}
\end{table}

\printfigures

\newpage
\printtables

\end{document}